\documentclass[%
reprint,
nobibnotes,
amsmath,amssymb,
aps,
prb,
floatfix,
]{revtex4-2}

\usepackage[utf8]{inputenc}
\usepackage[T1]{fontenc}
\usepackage{graphicx}
\usepackage{dcolumn}
\usepackage{bm}
\usepackage{xcolor}
\usepackage{amsmath}
\usepackage{mathtools}
\usepackage{braket}
\usepackage[caption=false, position=top]{subfig}
\usepackage{lipsum}
\usepackage{hyperref}
\usepackage{booktabs}
\usepackage{array}
\usepackage{multirow}
\usepackage{longtable}
\usepackage{siunitx}
\DeclareSIUnit{\angstrom}{\text{\AA}}
\usepackage{physics}

\makeatletter
\def\nobreakhline{%
  \noalign{\ifnum0=`}\fi
    \penalty\@M
    \futurelet\@let@token\LT@@nobreakhline}
\def\LT@@nobreakhline{%
  \ifx\@let@token\hline
    \global\let\@gtempa\@gobble
    \gdef\LT@sep{\penalty\@M\vskip\doublerulesep}
  \else
    \global\let\@gtempa\@empty
    \gdef\LT@sep{\penalty\@M\vskip-\arrayrulewidth}
  \fi
  \ifnum0=`{\fi}%
  \multispan\LT@cols
     \unskip\leaders\hrule\@height\arrayrulewidth\hfill\cr
  \noalign{\LT@sep}%
  \multispan\LT@cols
     \unskip\leaders\hrule\@height\arrayrulewidth\hfill\cr
  \noalign{\penalty\@M}%
  \@gtempa}
\makeatother

\newcolumntype{P}[1]{>{\centering\arraybackslash}p{#1}}
\newcolumntype{M}[1]{>{\centering\arraybackslash}m{#1}}

\newcommand{\eps}{\varepsilon}

\newcommand{\wt}{\widetilde}
\renewcommand{\r}{{\vb{r}}}
\newcommand{\q}{{\vb{q}}}
\newcommand{\G}{{\vb{G}}}

\newcommand{\p}{\partial}

\begin{document}
\preprint{APS/123-QED}

\title{Quantitative prediction of excitons in lattice-mismatched van der Waals heterostructures}

\author{Jakob Kjærulff Svaneborg$^{1}$}
\email{jakoks@dtu.dk}
\author{Mikkel Ohm Sauer$^{1}$}
\author{Amalie Helena Svaneborg$^{1}$}
\author{Kristian Sommer Thygesen$^{1}$}

\affiliation{$^1$CAMD, Department of
Physics, Technical University of Denmark, DK - 2800 Kongens Lyngby,
Denmark}

\date{\today}

\begin{abstract}
Accurate modeling of dielectric screening in van der Waals (vdW) heterostructures is essential for predicting photonic and optoelectronic properties - yet conventional first-principles methods are often hindered by incommensurate lattices and prohibitive computational costs. In this work, we introduce the microscopic Quantum Electrostatic Heterostructure (mQEH) method. mQEH employs a hierarchical and systematically improvable basis set to describe potentials and induced densities, eliminating the need for arbitrary geometric cutoffs and ensuring accurate screening descriptions at all length scales. The mQEH method is combined with a layer projected Bethe-Salpeter Equation (BSE) to enable calculations of optical spectra of experimentally relevant lattice-mismatched vdW heterostructures. Applying the mQEH-BSE framework to a series of transition-metal dichalcogenide (TMD) heterobilayers, we obtain absorption spectra and momentum-indirect exciton energies in excellent agreement with experiment. The framework provides a computationally efficient route to predictive modeling and design of vdW heterostructures with tailored optical properties.
 
\end{abstract}

\maketitle

\section{Introduction}

Van der Waals heterostructures composed of stacked layers of atomically thin two-dimensional (2D) crystals\cite{geim2013van}, offer an unprecedented degree of control over material properties at the atomic scale. By combining layers with diverse chemical compositions, crystal structures or symmetries, it is possible to control electronic, optical, and topological properties of the resulting stack\cite{ren20262d,de2025roadmap}. This concept has already enabled applications ranging from light-emitting diodes\cite{withers2015light} and photodetectors\cite{long2019progress} to excitonic transistors\cite{ciarrocchi2022excitonic} and moiré-engineered correlated electron phases\cite{cao2018unconventional,wang2020correlated}. However, the vast space of possible stack combinations - defined by layer types, stacking order, and twist angle - renders traditional trial-and-error approaches impractical, calling instead for rational design strategies based on computational screening to identify the most promising candidates. 

Unfortunately, the complex nature of the vdW heterointerface makes a quantitative description from first principles highly challenging. In particular, weak interlayer bonding permits lattice mismatched structures, often leading to unit cells containing hundreds or even thousands of atoms\cite{bistritzer2011moire}. At the same time, weak dielectric screening in the 2D limit enhances many-body effects, calling for descriptions beyond the mean field level of density functional theory (DFT)\cite{cudazzo2011dielectric,thygesen2017calculating}. Finally, the intimate coupling between band alignment, charge transfer, and interfacial dipoles, makes a self-consistent solution of the electronic structure strictly necessary.

A natural way to address the challenges described above is to exploit the weak interlayer coupling and formulate theories in which the individual layers are treated accurately while their mutual interactions are treated approximately.

Examples of this strategy include the quantum electrostatic heterostructure (QEH) model\cite{andersen2015dielectric} and the multilayer Rytova–Keldysh (MLRK) Bethe–Salpeter equation (BSE)\cite{sauer2022exciton}. The QEH model obtains a coarse-grained dielectric function of a stack by coupling the monopole and dipole components of the dielectric functions of the isolated layers through the Coulomb interaction, thereby neglecting interlayer hybridization altogether. The MLRK method evaluates the BSE matrix elements of the electron–hole interaction in a mixed in-plane/out-of-plane basis projected onto the individual layers, using a Keldysh model for the screened interaction. While these methods are efficient, they are still approximate in nature and lack the quantitative predictive power of a true first principles approach. 

In this work, we develop a unified computational framework to predict correlated excitations in general vdW heterostructures. The foundation of the framework is a QEH-inspired model for dielectric screening that extends far beyond previous models by accounting for microscopically varying densities and potentials crucial in \emph{ab initio} many-body calculations. The crux of this model (referred to as mQEH) is the use of a hierarchical dielectric basis set to represent induced densities. By expanding electron-hole (e-h) pair densities in terms of this dielectric basis, the mQEH can be leveraged to construct the BSE Hamiltonian in a layer-resolved manner. 

Combining the mQEH-BSE framework with our recently developed LAPS method for quasiparticle (QP) band structures\cite{leon2025laps}, we obtain the excitonic spectrum of the four monolayer transition-metal dichalcogenides (TMDs) MoS$_2$, MoSe$_2$, WS$_2$, and WSe$_2$, homobilayer MoS$_2$, and all six TMD heterobilayer combinations. Across all systems, we find excellent agreement with experiment, with errors on the bilayer exciton energies generally below 100 meV. We analyze the spin, k-space and real space structure of the lowest interlayer excitons, and explain the twist angle-dependence of selected excitons. Our work represents, to our knowledge, the first systematic, quantitative comparison of first-principles calculations with experiments for excitons in vdW heterostructures and 
paves the way for predictive modeling of the electronic and optical properties of experimentally relevant semiconducting vdW materials.

\section{Theory}
\subsection{Dielectric screening with mQEH}
\label{sec:theory-mqeh}
A fundamental ingredient in both QP band structure and exciton calculations, is the dielectric response function. Within the random-phase approximation (RPA), the density-response function $\chi$ satisfies the Dyson equation
\begin{equation}
\label{eq:main-dyson}
    \chi = \chi_0 + \chi_0 V \chi,
\end{equation}
where $\chi_0$ is the independent-particle response function obtained from Kohn--Sham DFT, and $V$ is the Coulomb interaction.

Assuming a basis of layer-localized density functions, the problem of representing $\chi_0$ ($\chi$) and solving Eq. (\ref{eq:main-dyson}) can be divided into an intralayer and interlayer problem. Intralayer quantities are denoted by a tilde. Splitting the Coulomb interaction as $V = \wt{V} + V_I$ and denoting the density response function of the isolated layer $\ell$ by $\wt{\chi}^\ell$, the Dyson equation can be cast in the form
\begin{equation}
\label{eq:dyson-intralayer}
\wt{\chi}^\ell = \wt{\chi}^\ell_0 + \wt{\chi}^\ell_0 \wt{V} \wt{\chi}^\ell,
\end{equation}
\begin{equation}
\label{eq:dyson-interlayer}
\chi = \wt{\chi} + \wt{\chi} V_I \chi,
\end{equation}
where 
\begin{equation}
\wt{\chi}(\r, \r') =  \sum_\ell \wt{\chi}^\ell(\r, \r')    
\end{equation}
is the response function of the heterostructure in the absence of any interlayer interactions. The intralayer Dyson equation \eqref{eq:dyson-intralayer} is readily solved, e.g. in a plane wave representation, using standard first-principles RPA implementations. 

Next, we address the crucial question of choosing an optimal basis for representing intralayer and interlayer quantities and solving the interlayer Dyson equation \eqref{eq:dyson-interlayer}. The QEH \cite{andersen2015dielectric, gjerding2020efficient} and QEH-inspired models \cite{macheda2024ab} achieve this by defining a minimal layer-dependent monopole/dipole basis with a parameter $d_\ell$ which acts as an effective thickness of each monolayer.
This approach, while efficient, has two central weaknesses: it does not permit an extension of the basis to accurately describe microscopic variations of the induced fields, and it only works for mirror-symmetric $(z \rightarrow -z)$ monolayers, which precludes the description of Janus materials and other monolayers which lack this symmetry. 

We overcome these limitations, by introducing a hierarchical basis of dielectric eigen functions in each layer. The basis is designed to cleanly separate intra- and interlayer interactions while capturing microscopic screening effects, which are crucial for many-body perturbation theory calculations. It is also computationally efficient: for any monolayer, independent of symmetry, the dielectric basis can be systematically expanded toward completeness in order of increasing relevance to screening. We shall refer to the resulting model as the microscopic QEH (mQEH) model.

Since electrostatic potentials are inherently long-ranged, while induced charge densities are strongly localized within individual layers, we represent these quantities using distinct basis sets.
The potential basis functions,  $\phi_{\alpha}^\ell$, for layer $\ell$ are chosen as the eigen functions of the operator $V\wt{\chi}^\ell(\omega=0)$, given by  
\begin{equation}
\label{eq:Vchi}
    (V\wt{\chi}^\ell)(\q, z, z') = \int \dd z'' V(q, z, z'') \wt{\chi}^\ell(\q, \omega=0, z'', z'), 
\end{equation}
where $V(q, z,z') = \frac{2\pi}{q} e^{-q |z-z'|}$ is 
the 2D Fourier transform of $\frac{1}{|\r - \r'|}$, and $q = |\q|$. We obtain the basis functions from the static ($\omega=0$) response function, because their $\omega$-dependence is weak anyway. In principle, one can use $\omega$-dependent eigen functions, but it is not strictly necessary as the $\omega=0$ basis is complete and thus calculations can always be converged by including more basis functions. 

\begin{figure}
    \centering
    \includegraphics{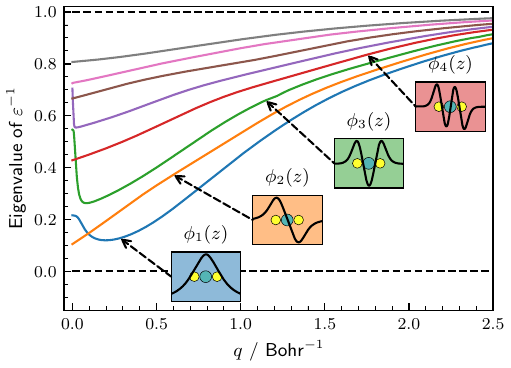}
    \caption{Eigenvalues of $\tilde{\eps}^{-1} = 1 + V \wt{\chi}$ for MoS$_2$. The spectrum lies entirely in the interval $[0,1]$; an eigenvalue of 0 corresponds to perfect screening ($V_\text{ind} = - V_\text{ext}$) while an eigenvalue of 1 corresponds to no screening ($V_\text{ind} = 0$).
    The insets show the spatial profile of some of the eigen functions evaluated at different in-plane momenta $q$. }
    \label{fig:MoS2-dielectric-eigenfunctions}
\end{figure}

The motivation for using the eigen functions of $V\wt{\chi}^\ell$ to represent potentials in layer $\ell$, is the following: 
the total (screened) potential may be written as the sum of the external and induced potentials, $V_\text{tot} = V_\text{ext} + V_\text{ind}$. 
Within linear response theory, the induced potential is given by $V_\text{ind} = (V\chi)V_\text{ext}$. Compared to the external potential, the induced potential is generally smaller in magnitude and opposite in sign, reflecting the fact that the total strength of the interaction is reduced by electronic screening. 
Therefore, the static eigenvalues of $V\wt{\chi}^\ell$ are negative, and lie inside the interval $[-1, 0]$. 
To efficiently describe the most important screening channels in the smallest possible basis, we may therefore retain the $N_\ell$ basis functions corresponding to the largest magnitude eigenvalues of $V\wt{\chi}^\ell(\omega=0)$.
The number $N_\ell$ of basis functions retained for any given layer can be chosen according to the desired accuracy of the calculation and the available computational resources.
In general, we have found that including more than 6-8 basis functions per layer results in negligible changes to the calculated physical properties.
The spectrum of $\tilde{\eps}^{-1}(q) = 1 + V(q)\wt{\chi}(q)$ for MoS${}_2$ is shown in Fig.~\ref{fig:MoS2-dielectric-eigenfunctions} along with the spatial profiles of a selection of the potential basis functions $\phi^\ell_\alpha(q,z)$. We note that a similar basis has been previously used by \cite{andersen2012spatially} to describe plasmon modes in metallic nano-films.

As $V\wt{\chi}^\ell$ is not hermitian, the eigen functions $\phi^\ell_\alpha$ are not orthogonal. 
Instead, we may define a dual basis $\rho^\ell_n$ by the condition
\begin{equation}
\label{eq:orthonormality-rho-phi}
    \braket{\rho^\ell_\alpha(q)}{\phi^\ell_\beta(q)} = \delta_{\alpha\beta}.
\end{equation}

It can be shown that the $\phi^\ell_\alpha$ and $\rho^\ell_\alpha$ are related by the Poisson equation\cite{andersen2012spatially}, which in the mixed coordinate representation reads
\begin{equation}
    (\p_z^2 - q^2) \phi^\ell_\alpha(q,z) = - 4\pi \rho^\ell_\alpha(q,z),
\end{equation}
from which it may also be shown that $ \wt{\chi}^\ell \ket{\phi^\ell_\alpha} = \wt{\chi}^\ell_\alpha \ket{\rho^\ell_\alpha}$, where $\wt{\chi}^\ell_\alpha$ is a scalar. In other words, the functions $\rho^\ell_\alpha$ are precisely the charge densities induced in the free-standing monolayer $\ell$ by a static potential with the spatial profile $\phi^\ell_\alpha(z)$. While the potential eigen functions $\phi^\ell_\alpha$ may be delocalized and extend over the entire heterostructure, the densities $\rho^\ell_\alpha$ are exponentially localized on each layer. This property, along with Eq.~\eqref{eq:orthonormality-rho-phi} allows us to use the $\rho^\ell_n$ as basis functions for the electron density and as local projection functions for potentials. In particular, this permits us to construct the local representations of the Coulomb kernel and density response functions required by Eqs.~(\ref{eq:dyson-intralayer}--\ref{eq:dyson-interlayer}) without the need to introduce any parameters, such as the center and spatial extent of each layer, as required by earlier methods \cite{macheda2024ab, andersen2015dielectric, gjerding2020efficient}.
More detail on the representation of operators in this basis may be found in Appendix \ref{appendix:mqeh-basis}.

Once we have solved Eq.~(\ref{eq:dyson-interlayer}) and found the density response function, $\chi$, for the entire heterostructure in the mQEH basis, we can directly construct the dielectric function $\eps^{-1} = 1 + V \chi$ and the screened interaction $W = \eps^{-1} V$, which is represented via its matrix elements in the density basis,
\begin{equation}
\label{eq:W-mqeh}
    W_{\alpha \ell, \beta \ell'}(\q, \omega)
    = \int \dd z \dd z' 
    \rho^\ell_\alpha(\q, z)^* W(\q, \omega, z, z')
    \rho^{\ell'}_\beta(\q, z').
\end{equation}
These matrix elements have a direct physical interpretation as the interaction energy of the screened interaction between the charge distributions $\ket{\rho^\ell_\alpha}$ and $\ket{\rho^{\ell'}_\beta}$.

We stress that local field effects (LFE) within each monolayer are fully accounted for when solving Eq.~\eqref{eq:dyson-intralayer}. Only interlayer LFE, describing scattering between different in-plane momenta by charge fluctuations in a neighboring layer, are neglected, see Appendix \ref{appendix:mixed-space}. A consequence of this, in addition to the neglect of interlayer hybridization is that the mQEH model cannot distinguish between different stacking orders of the heterostructure. However, it enables us to describe the dielectric response of heterostructures with non-commensurate or large Moiré supercells at no additional computational cost. In practice, this is a great trade-off because, as previously mentioned, dielectric properties are insensitive to the small changes in electronic structure caused by the detailed stacking configuration.

\begin{figure}
    \centering
    \includegraphics{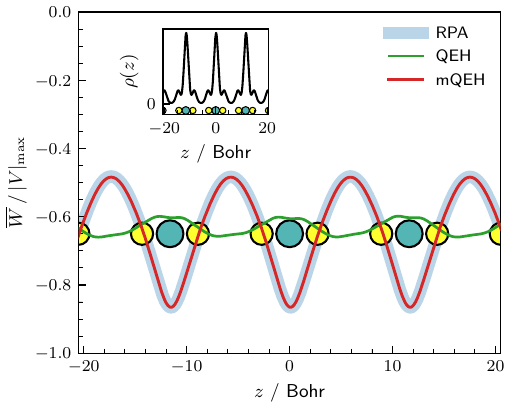}
    \caption{Spatial profile of the screened interaction $\overline{W} = W - V$ in bulk MoS$_2$ induced by an external potential $V_\text{ext}(q, z, \omega = 0)$. The external potential is generated by the charge density shown in the inset, and has in-plane variation $e^{i\vb{q}\cdot \r_\parallel}$ with $|\vb{q}| = 0.3$ Bohr$^{-1}$ in the $\Gamma-M$-direction. The charge density corresponds to the $z$-profile of the valence-band wavefunction in a MoS${}_2$ monolayer: $\rho(z) = \int |\psi_\text{VBM, ML}(\vb{r})|^2  \dd \vb{r}_\parallel$.  The mQEH response is computed with 8 basis functions. For the QEH and mQEH responses, the bulk structure is simulated by a heterostructure containing 10 layers. Only the response in the central layers in the stack is shown. The RPA calculation is performed with full periodic boundary conditions. 
    }
    \label{fig:mQEH_realspace}
\end{figure}

To illustrate the performance of the mQEH model, we consider its ability to reproduce the exact RPA screened Coulomb interaction inside a layered material. 
Fig. \ref{fig:mQEH_realspace} shows the spatial out-of-plane profile of the screened interaction $\overline{W} = W - V$ in bulk MoS$_2$ induced by the potential associated with a charge density of the form $\varrho(z,\mathbf q_{||})=\varrho(z)e^{i\vb{q}\cdot \r_\parallel}$, where $\varrho(z)$ (see inset) represents the density of a valence band state. Using only eight dielectric basis functions, mQEH reproduces the exact RPA results perfectly. In contrast, the QEH result does not capture the microscopic variations in the screened potential, although it yields a reasonable averaged value.

\subsection{Excitons with mQEH-BSE}
Exciton excitation energies may be obtained by solving the Bethe--Salpeter equation (BSE).
The main computational bottleneck in BSE calculations for vdWHs with large unit cells is the evaluation of the required matrix elements of the screened Coulomb interaction. This is typically obtained in the RPA, and requires a converged dielectric response function $\eps^{-1}$ with screening from a large number of high-energy electronic states. Below we describe a method to simplify the calculation of the screened Coulomb matrix elements by projecting electron-hole pair densities into the mQEH dielectric basis functions. We refer to the method as mQEH-BSE.

The BSE can be expressed as an eigenvalue problem for a two-particle Hamiltonian $\mathcal{H}(\q)$. 
In the electron-hole basis $\ket{1, 2}$, whose elements consist of products of single-particle orbitals of the form
\begin{equation}
\label{eq:pair-density-realspace}
    \braket{\sigma \r, \sigma'\r'}{1, 2} =  \psi_{n_1, \vb{k_1}}^*(\sigma, \r)\psi_{n_2, \vb{k_2}}(\sigma', \r'),
\end{equation}
the matrix elements of $\mathcal{H}(\q)$ are
\begin{equation}
    \label{eq:BSE}
\begin{split}
\mel{1,2}{\mathcal{H(\q)}}{3, 4} 
& = (\eps_{2} - \eps_{1} ) \delta_{1,3} \delta_{2, 4}\\
& - \left(f_{2} - f_{1}\right) \mel{1, 2}{K(\q)}{3,4}.
\end{split}
\end{equation}
Here, $\eps_i$ and $f_i$ are the energy and occupation number, respectively, of the single-particle state $\ket{i}$. 
We note that all our calculations include spin-orbit coupling (SOC) except where stated otherwise. The single-particle states are therefore spinors, and inner products in spin-space are implicitly assumed in all matrix element expressions. As such, we will omit the $\sigma$ of Eq. \eqref{eq:pair-density-realspace}.
Conservation of crystal momentum requires that $\vb{k_2} = \vb{k_1} + \q $ and $\vb{k_4} = \vb{k_3} + \q$.
$K(\q)$ is the BSE kernel, which is given by 

\begin{equation}
\label{eq:bse-kernel-reciprocalspace}
\begin{split}
   \mel{1, 2}{K(\q)}{3,4} =& \frac{1}{\Omega} \sum_{\G \G'} \\
     \Bigg[
    &[\varrho^1_2 (\G)]^* V_{\G}(\q) \delta_{\G \G'}  \varrho^3_4 (\G) \\
    - &[\varrho^1_3 (\G)]^* W_{\G \G'}(\vb{k}_3 - \vb{k}_1, \omega=0)  \varrho^2_4 (\G')
    \Bigg],
\end{split}
\end{equation}
where the so-called \textit{pair densities}
\begin{equation}
\label{eq:pair-density}
\varrho^i_j(\G) = \int \dd \r \psi_i^*(\r) e^{-i (\G+ \vb{k}_j - \vb{k}_i) \r} \psi_j(\r),
\end{equation}
are the Fourier transform of $\braket{\r, \r}{i, j}$.
To find the lowest-order excitons, it is typically enough to include only a few bands closest to the Fermi level in the BSE Hamiltonian. 
However, as previously mentioned, to converge the screened interaction $W$, which is required to construct the BSE Hamiltonian, one must include a large number of high-energy excitations in the calculation of $\eps^{-1}$.
For structures with large unit cells, this becomes unfeasible. 

To overcome this problem, we calculate $W$ using the mQEH model. 
To construct the required matrix elements of $W$ from the mQEH representation Eq.~(\ref{eq:W-mqeh}), we first cast the pair densities in the mixed-space basis representation,
\begin{equation}
    \varrho^i_j(\G_\parallel, z) 
    = \int \dd \r_\parallel \psi_i^*(\r_\parallel, z)
    e^{-i (\G_\parallel+ \vb{k}_j - \vb{k}_i) \r_\parallel} 
    \psi_j(\r_\parallel, z).
\end{equation}
For every (fixed) $\G_\parallel$, we project the function $z \rightarrow \varrho^i_j(\G_\parallel, z)$ into the space spanned by the mQEH density basis functions $\rho^\ell_\alpha$,
\begin{equation}
\label{eq:pairdensity-in-mqeh-basis}
    \varrho^i_j(\G_\parallel, z) = \sum_{\ell\alpha} C^{i \ell}_{j \alpha}(\G_\parallel) \rho^\ell_\alpha(\G_\parallel + \vb{k}_j - \vb{k}_i, z)
\end{equation}

Inserting Eq.~(\ref{eq:pairdensity-in-mqeh-basis}) in Eq.~(\ref{eq:bse-kernel-reciprocalspace}), the final expression for the direct (screened) part of the kernel becomes
\begin{equation}
\begin{split}
    &\mel{1, 2}{W}{3, 4} \\
    &=\sum_{\alpha \ell,\beta \ell', \G_\parallel}
    [C^{1 \ell}_{3 \alpha}(\G_\parallel)]^* C^{2 \ell'}_{4\beta}(\G_\parallel)
    W_{\alpha \ell \beta \ell'} (\vb{k}_3 - \vb{k}_1 + \G_\parallel).
\end{split}
\end{equation}

\begin{figure}
    \centering
    \includegraphics{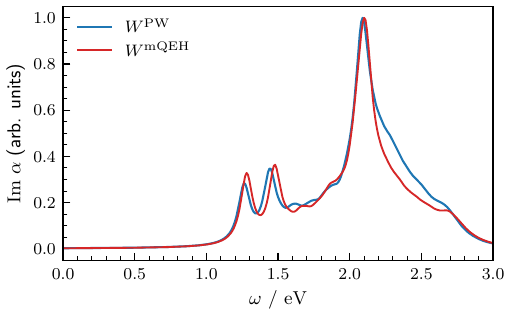}
    \caption{BSE@PBE absorption spectrum of a lattice-matched MoS$_2$-WS$_2$ bilayer, calculated without spin-orbit coupling. This structure contains only 6 atoms per unit cell, making full plane-wave BSE calculations feasible. Such a calculation is compared here to the spectrum computed with the mQEH-BSE method. Both peak positions and heights are in excellent agreement between the two methods.}
    \label{fig:bse-mqeh-gpaw}
\end{figure}

Fig. \ref{fig:bse-mqeh-gpaw} compares the excitation spectrum obtained with mQEH-BSE to a full plane wave BSE calculation for lattice matched MoS$_2$/WS$_2$. This system has only six atoms in a unit cell, which makes it possible to perform well converged conventional BSE calculations. There is an excellent agreement between the two spectra with less than 50 meV difference in the position of the main exciton peaks. Note that the calculation is performed with the PBE band structure and without SOC. The spectra are therefore not quantitatively correct, but sufficient for the purpose of technical benchmarking. 

Finally, we mention that the mQEH-BSE calculation converges faster with k-points than the conventional BSE method, see Figure \ref{fig:bse-convergence}. This is because the dielectric building blocks are already produced with a very fine k-point mesh whereas the conventional BSE implementation evaluates the screening at the same k-point grid as used to setup the BSE Hamiltonian.

\subsection{Quasiparticle energies with LAPS}
To obtain a quantitatively correct description of excitons in a heterostructure, it is crucial that the underlying  QP energies are accurate. In particular, the band gaps of the involved monolayers and the relative band alignment between neighboring monolayers is crucial. Because the band alignment is intricately tied to charge transfer and interfacial dipoles, a self-consistent approach is required.   It is well known that (semilocal) DFT tends to underestimate QP band gaps and yield wrong charge transfer and dipoles at hetero-interfaces. If the underlying DFT ground state calculation does not capture the correct band alignment, the interfacial charge transfer, ground state electron density, and single-particle wave functions will also be wrong. Subsequent perturbative approaches based on such DFT ground states, like the popular G$_0$W$_0$ scheme, will then inherit these flaws. On the other hand, self-consistent GW methods are generally computationally infeasible for structures containing hundreds of atoms in the unit cell.

\begin{table}
\center
\begin{ruledtabular}
\begin{tabular}{lp{5pt}ccc}
  \\[-5pt]
   Material/Method                && DFT-PBE & G$_0$W$_0$ & G$_0$W$_0\Gamma_0$ 
  \\[5pt]
  \hline\\[-3pt] 
   MoS$_2$  VBM        &&-5.96 &-6.16 &-5.79            \\ 
   MoS$_2$  CBM        &&-4.28 &-3.66 &-3.32            \\ 
   MoS$_2$  Band gap        && 1.68 & 2.50 &2.47 
   \\ 
   \\[-3pt]
   MoSe$_2$ VBM        &&-5.30 &-5.51 &-5.14            \\ 
   MoSe$_2$ CBM        &&-3.85 &-3.33 &-2.94            \\ 
   MoSe$_2$ Band gap        && 1.45 & 2.19 & 2.20            \\ 
   \\[-3pt]
   WS$_2$   VBM        &&-5.71 &-6.21 &-5.80            \\ 
   WS$_2$   CBM        &&-3.90 &-3.48 &-3.14            \\ 
   WS$_2$   Band gap        && 1.81 & 2.73 &2.66 
   \\ 
   \\[-3pt]
   WSe$_2$  VBM        &&-5.06 &-5.57 &-5.17 
   \\ 
   WSe$_2$  CBM        &&-3.51 &-3.17 &-2.86 
   \\ 
   WSe$_2$  Band gap        && 1.55 & 2.40 &2.31              
   \\[3pt]
 \end{tabular}
 \end{ruledtabular}
 \caption{Energies of the valence band maximum (VBM) and conduction band minimum (CBM) relative to vacuum and the band gap of the four monolayer TMDs calculated at three levels of theory. Spin orbit coupling is not included. The values are used to define the scissors corrections in the LAPS method. 
 } 
  \label{tab:GWmonolayers}
 \end{table}
 
Here we address the problem using the self-consistent layer projected scissors operator (LAPS) methodology~\cite{leon2025laps}.
In this method, a LAPS operator $\Sigma_\text{LAPS}$,
\begin{equation}
    \label{eq:LAPS-operator}
    \Sigma_\text{LAPS}
    = \sum_\ell^\text{layers} \left[
    \Delta \eps_{v, \ell} P_\ell^\text{occ}
    + \Delta \eps_{c, \ell} \left(1 - P_\ell^\text{occ}\right)
    \right],
\end{equation}
where $P_\ell^\text{occ}$ is the projector into the occupied electronic subspace of layer $\ell$ and $\Delta \eps_{v/c, \ell}$ are so-called scissors shifts, are added to the Kohn--Sham Hamiltonian. The projection operators depend on the density matrix and are therefore updated during the self-consistent DFT cycle.   
The values of the scissors shifts for each layer are input parameters to the LAPS operator. The main contribution to the scissors shifts of a given layer is set by the condition that LAPS should reproduce the target values for the valence and conduction band edges for the isolated layer. In this work the target values are obtained from vertex-corrected G$_0$W$_0$ following the G$_0$W$_0\Gamma_0$ method, where the vertex is taken as the renormalised adiabatic LDA xc-kernel\cite{schmidt2017simple}. We stress that the non-selfconsistent nature of the G$_0$W$_0\Gamma_0$ is not a problem here, because it is only used to obtain the band edge positions of the isolated monolayers where interfacial charge transfer obviously does not occur. The calculated band edge energies of the four TMD monolayers are listed in Table \ref{tab:GWmonolayers}. In addition to the difference between the DFT and target band edge energies of the freestanding monolayer, the scissors shifts include a smaller correction for image charge renormalization of the band edges arising due to the additional screening from the other layer of the heterostructure. This correction, which is on the order $\pm 0.1$ eV for the occupied/unoccupied bands is calculated with the G$\Delta$W-method\cite{winther2017band}.  

We refer to \cite{leon2025laps} for a precise definition of the projectors $P^\text{occ}_\ell$, validation of the method, and for further discussions on the determination of the scissors shifts. Adding $\Sigma_\text{LAPS}$ to the Kohn-Sham Hamiltonian ensures that interlayer charge transfer is taken into account self-consistently, and that the band gaps and band alignment equals the target GW values in the zero-hybridization limit. 

Full computational details for the ground-state, mQEH, and BSE calculations are provided in Appendix~\ref{appendix:computational-details}.


\section{Results}

\subsection{Monolayers}
Although the mQEH-BSE method is designed for vdW heterostructures, we first apply it to the four freestanding TMD monolayers. This provides a useful baseline for assessing the description of intralayer excitons before turning to the more complex case of TMD heterobilayers.

\begin{figure*}[ht]
    \centering
    \includegraphics{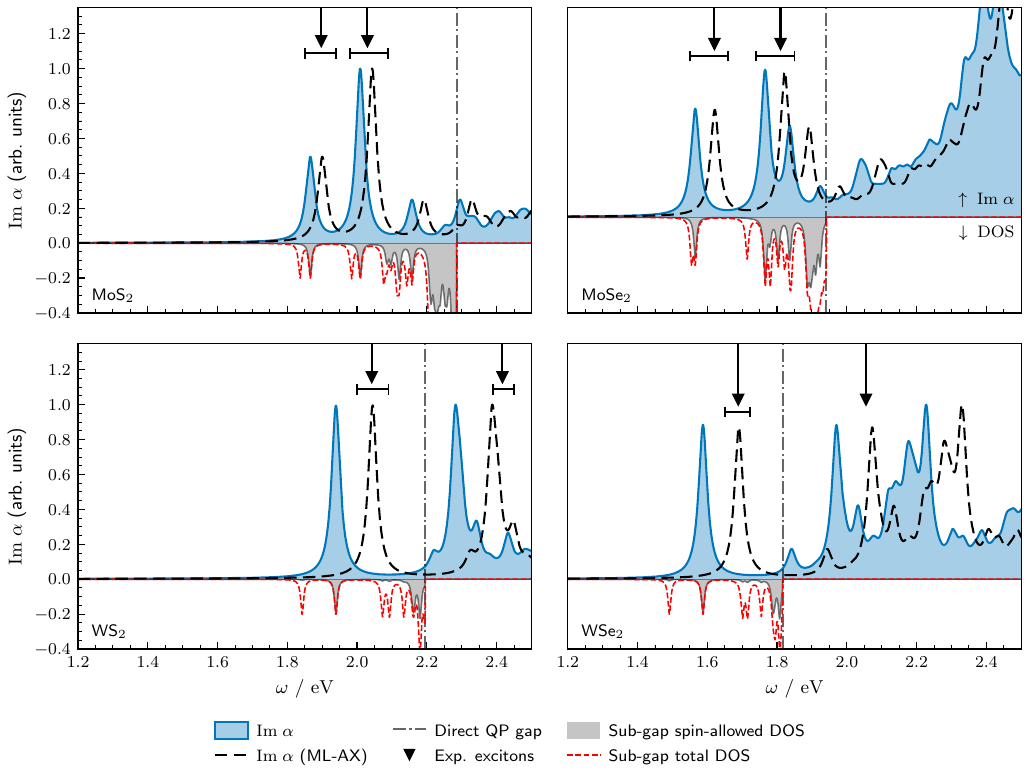}
    \caption{Absorption spectra and (negated) exciton density of states (DOS) for the four TMD monolayers considered in this work. The downward arrows show experimentally determined A and B exciton energies with the variation across different experiments indicated by horizontal bars (experimental values are given in Table \ref{TableEXP_ML}). The blue-shaded spectra correspond to G$_0$W$_0 \Gamma_0$ band gaps. The black dashed spectra are obtained after including a band gap correction in the LAPS operator.}
    \label{fig:monolayer-spectra}
\end{figure*}

Figure \ref{fig:monolayer-spectra} shows the absorption spectra calculated with mQEH-BSE. The blue-shaded spectrum is obtained using band-edge target values in the LAPS operator corresponding to G$_0$W$_0\Gamma_0$\cite{schmidt2017simple}. It therefore represents a fully \emph{ab initio} result with no free parameters. The experimental positions of the A and B excitons are indicated by downward arrows, with horizontal bars showing the spread across different experimental reports; see Table \ref{TableEXP_ML}. For the Mo-based TMDs, the two lowest absorption peaks in the mQEH-BSE spectra, corresponding to the A and B excitons, fall within the experimental range. In contrast, for the W-based TMDs, the calculated peaks are underestimated by approximately $150$ meV and lie below the experimental range.

It is tempting to ascribe the underestimation of the intralayer exciton energies to inaccuracies in the G$_0$W$_0\Gamma_0$ band gaps. Indeed, the band gaps obtained with G$_0$W$_0\Gamma_0$@PBE are very close to those predicted by G$_0$W$_0$@PBE, which is known to systematically underestimate quasiparticle band gaps in semiconductors and insulators \cite{shishkin2007self,huser2013quasiparticle}. This deficiency can be remedied by including an additional correction in the LAPS operator given by the difference between the predicted and the experimentally measured A exciton energy. The resulting absorption spectra are shown as black dashed lines in Fig.~\ref{fig:monolayer-spectra}. As expected, the A exciton peaks in the corrected spectra are then brought into excellent agreement with experiment. 
Notably, the B exciton peaks in the corrected spectrum also match experiment closely.

Since the band-edge corrections were derived directly from the monolayer spectra, their application does not, by itself, provide additional insight for the isolated monolayers. 
However, these corrections can  optionally be carried over to heterostructure calculations.
Including these corrections in the LAPS operator  ensures that the monolayer band gaps and intralayer exciton energies are recovered exactly in the dissociation limit, i.e. at infinite layer separation.
Whether this empirical refinement  improves the agreement with experiment for a given heterostructure is a separate question, which we address below.
For isolated monolayers, only the total band-gap correction affects the exciton energies. In heterostructures, however, the partitioning of this correction between the valence- and conduction-band edges can become important, as it determines the relative band alignment between the layers. Below we always partition the correction evenly between the occupied and unoccupied band manifolds. We refer to this correction as the monolayer A-exciton correction (ML-AX correction).

To illustrate the energetic distribution of all excited states, independent of their oscillator strengths, we show the (negated) density of states (DOS) below the absorption spectrum. Only states below the QP band gap are included. The total DOS is shown as a red dashed curve, while the optically active singlet states are indicated by the gray-shaded area. For all four monolayers, the lowest excitation is found to be a dark triplet state, in agreement with previous first-principles studies \cite{deilmann2017dark}.

\subsection{Homobilayer MoS$_2$}
As a natural next step toward heterobilayers, we consider homobilayer MoS$_2$ in the most stable 2H (AB$_6$) stacking configuration\cite{pakdel2024high}. Whereas the four TMD monolayers all have direct band gaps at the K point, their homobilayer counterparts have indirect band gaps. This change arises from interlayer hybridization, which shifts the valence-band maximum upward at the $\Gamma$ point. The interlayer hybridization and the resulting transition from a direct to an indirect band gap therefore represent a step up in complexity relative to the monolayers. Nevertheless, homobilayers remain simpler than heterobilayers: their commensurate lattices give rise to a primitive unit cell containing only six atoms, and the identical electronegativities of the two layers — and even inversion symmetry in the stacking considered here — suppress interlayer dipoles and charge transfer.

\begin{figure*}
       \centering
    \includegraphics{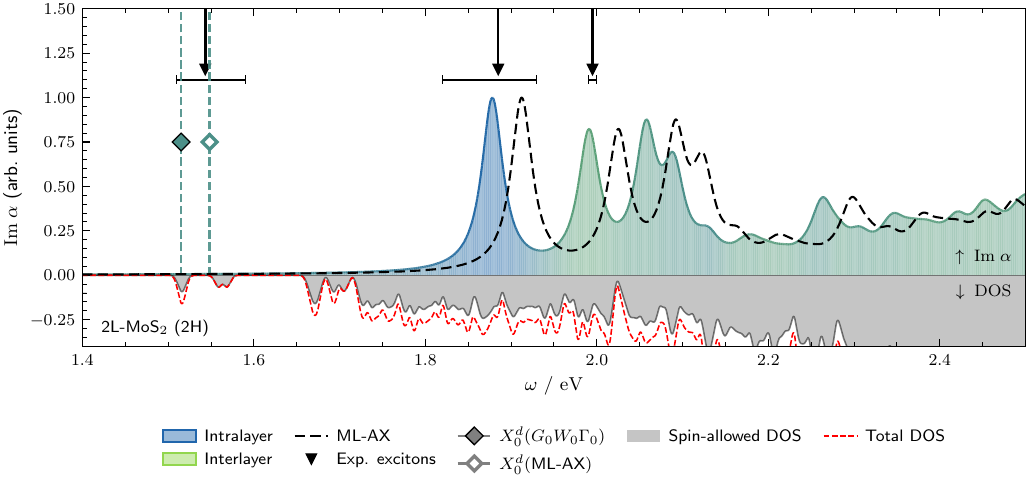}
    \caption{Absorption spectrum of homobilayer H-MoS${}_2$ in the 2H stacking configuration. The downward arrows show experimentally determined exciton energies (see values in Table \ref{TableEXP_2L_MoS2}). Blue and green colors represent the degree of intralayer and interlayer character, respectively. The (negated) exciton DOS is shown below the absorption spectrum.}
    \label{fig:MoS2-bilayer-spectra} 
\end{figure*}

Figure \ref{fig:MoS2-bilayer-spectra} shows the absorption spectrum calculated with mQEH-BSE. The color-shaded spectrum is obtained using a LAPS operator constructed from the G$_0$W$_0\Gamma_0$ band-edge values of freestanding monolayer MoS$_2$, corresponding to a fully \emph{ab initio} approach. The black dashed curve shows the spectrum obtained after adding the ML-AX corrections to the LAPS operator. We denote the lowest (optically dark) exciton by $X_0^d$. Its experimentally measured position is indicated by the first downward-pointing arrow. The measured position of the two lowest, optically active excitons (often referred to as the A and B excitons) are indicated by the second and third downward-pointing arrows. The experimental data are provided in Table \ref{TableEXP_2L_MoS2}. 

Both the fully \emph{ab initio} and the ML-AX corrected calculations yield a lowest absorption peak in agreement with the experimentally determined position of the A peak. 
 We note that low temperature measurements yield an A exciton energy of 1.93 eV in excellent agreement with the ML-AX corrected spectrum. 
The blue color of the absorption peak indicates that the A exciton has predominantly intralayer character. This can be understood from the weak interlayer hybridization of the direct K--K transitions that form the A exciton: the Coulomb energy gained by localizing the electron and hole in the same layer dominates over the kinetic energy associated with hopping of either carrier between the layers.

The blue-green color of the spectrum above the A peak indicates that higher-lying excitations have mixed intralayer/interlayer character. This behavior can be attributed to two effects. First, higher-lying excitations involve transitions away from the K point, where interlayer hybridization is generally stronger. Second, the joint density of states increases with energy, so excitons can generally be formed from many transitions with both intralayer and interlayer character. For example, the second peak in the absorption spectrum is of such mixed intralayer/interlayer character. This particular exciton, has previously been described as a hybrid between the intralayer B-exciton and a pure interlayer A exciton\cite{peimyoo2021electrical}. Our analysis of the excited state giving rise to the second absorption peak agrees with this picture. As can be seen, the energy of this exciton also agrees well with the experimentally reported values.

The absorption spectrum includes only excitons with vanishing center-of-mass momentum, i.e. excitons comprised of direct electron-hole transitions. To probe the energy of the lowest indirect, and thus momentum-dark, excitons, we perform mQEH-BSE calculations at a finite momentum corresponding to the indirect $\Gamma-\mathrm{K}$ band gap of the bilayer ($Q=1/3$). The lowest singlet exciton energy, $X_0^d$, is indicated by diamonds in Figure \ref{fig:MoS2-bilayer-spectra}. The open and filled diamonds correspond to the G$_0$W$_0$$\Gamma_0$ LAPS operator with and without the ML-AX correction, respectively. Both approaches yield exciton energies within the experimental range. The blue-green color of the diamonds indicate a mixed interlayer/intralayer character of the indirect exciton. The intralayer character is driven by Coulomb attraction, which favors the electron and hole to localize at the same layer. The interlayer character is driven by interlayer hopping of the hole around the $\Gamma$-point, which favors delocalization of the hole over both layers.  

The (negated) exciton DOS is shown below the absorption spectrum in Figure \ref{fig:MoS2-bilayer-spectra}. The DOS represents the total distribution of eigenvalues obtained by diagonalizing the $Q=0$ and $Q=1/3$ mQEH-BSE Hamiltonians without including the ML-AX correction. The contribution from singlet excitons is indicated by the grey shading.

As in the monolayers, the lowest exciton is a triplet, and therefore spin-dark, exciton, in agreement with previous findings\cite{deilmann2019finite}.

\subsection{Heterobilayers}
We now turn to the six TMD heterobilayers. The calculated absorption spectra are shown in Fig. \ref{fig:mqehbse-spectra}. Experimental values for the lowest-lying (typically dark) exciton, $X_0^d$, and the lowest bright exciton, $X_0^b$, are indicated by the two downward-pointing black arrows. The horizontal bars indicate the variation in exciton energies from independent reports and/or from different twist-angles considered in the same study. We explore the twist-angle dependence in more detail in the next section. All experimental exciton energies are listed in Table \ref{TableEXP}.

For the heterobilayers, the ML-AX correction does not lead to a systematic improvement in the agreement with experiment. Therefore, Figure \ref{fig:mqehbse-spectra} shows only the fully \emph{ab initio} results, obtained using LAPS operators whose band-edge energies are taken from G$_0$W$_0\Gamma_0$ calculations (see the target values in Table \ref{tab:GWmonolayers}).
Details on the computational bilayer structures are provided in Table \ref{tab:moire-cells}. 

 \begin{table}[htbp]
    \centering
    \begin{tabular}{lccc}
      \hline
      Material        & Atoms & Angle (deg) & Strain (\%) \\
      \hline
      MoS$_2$/MoSe$_2$           & 75 & 16.1 & 0.09 \\
      MoS$_2$/WS$_2$             & 6  & 60.0 & 0.06 \\
      MoS$_2$/WSe$_2$            & 75 & 16.1 & 0.08 \\
      MoSe$_2$/WS$_2$            & 75 & 43.9 & 0.06 \\
      MoSe$_2$/WSe$_2$           & 6  & 60.0 & 0.03 \\
      MoSe$_2$/WSe$_2$           & 42 & 21.7 & 0.01 \\
      WS$_2$/WSe$_2$             & 75 & 43.9 & 0.05 \\
      \hline
    \end{tabular}
    \caption{Moiré supercell parameters for the TMD heterobilayers. The last column shows the strain on the monolayers in the heterobilayer supercell. }
    \label{tab:moire-cells}
  \end{table}
  
Across all six bilayers, the first absorption peak agrees very well with the experimentally measured lowest-lying optically active exciton, $X_0^\mathrm{b}$, indicated by the second downward-pointing arrow. In all cases, these excitations arise from transitions near the K point of the Brillouin zone. In MoS$_2$/MoSe$_2$, MoSe$_2$/WS$_2$, and WS$_2$/WSe$_2$, the lowest absorption peak has purely intralayer character, as indicated by its magenta or blue color. In the remaining three systems, the lowest absorption peak has mixed intralayer/interlayer character. This mixed character arises from either the electron or the hole being delocalized over both layers (due to interlayer hybridization) with the opposite charge being localized in one layer.

In the previous paragraph, we discussed the lowest optically active exciton, $X_0^\mathrm{b}$. To identify the lowest-lying, possibly dark, exciton, $X_0^\mathrm{d}$, we now consider the exciton energies, independently of their oscillator strengths.  In addition, we solve the mQEH-BSE at a finite center-of-mass momentum, $Q$, corresponding to the indirect band gap in the band structure. The energy of the lowest singlet exciton is indicated by the dashed vertical line. A diamond symbol denotes a momentum-indirect lowest exciton, whereas a circle denotes a momentum-direct one. The two bilayers MoSe$_2$/WS$_2$ and WS$_2$/WSe$_2$ fall into the latter category. In these systems, the lowest exciton has zero momentum, yet a vanishing, or negligible, oscillator strength. In MoSe$_2$/WS$_2$, this is due to the pure interlayer character of the exciton, which leads to negligible overlap of the electron and hole wave functions. In 
WS$_2$/WSe$_2$, the lowest exciton has mixed intralayer/interlayer character, and the vanishing oscillator strength is a result of the symmetry of the single-particle states comprising it. In the remaining four systems, the lowest exciton is momentum indirect and has either purely interlayer character (MoS$_2$/MoSe$_2$ and MoS$_2$/WSe$_2$) or mixed interlayer/intralayer character (MoS$_2$/WS$_2$ and MoSe$_2$/WSe$_2$).

As for the lowest optically active exciton, we find excellent agreement with experiment for the lowest exciton, $X_0^\mathrm{d}$. This is evident in Fig. \ref{fig:mqehbse-spectra} from the good agreement between the vertical dashed lines and the first downward-pointing arrow. Indeed, with the exception of MoS$_2$/WSe$_2$, all calculated exciton energies lie within 0.1 eV of the average experimental value, represented by the arrow. For MoS$_2$/WSe$_2$, the agreement is less satisfactory, which we attribute to the pronounced twist-angle dependence discussed in the next section.


\subsection{Twist-angle dependence}
We next consider the influence of the twist-angle on the lowest excitons. Our experimental data set for the three systems MoS$_2$/WS$_2$, MoSe$_2$/WSe$_2$, and  WS$_2$/WSe$_2$, comprises exclusively lattice matched structures (angles of $<1^\circ$ or $60^\circ$), see Table \ref{TableEXP}. For the first two bilayers (S/S and Se/Se), we also used lattice matched model structures for our calculations. Consequently, the obtained exciton energies can be directly compared to experiments. In both cases the agreement is highly satisfactory with the predicted exciton energies falling within 50 meV of the experimental range for both $X_0^{\mathrm{d}}$ and $X_0^{\mathrm{b}}$. 

\begin{figure*}[htp]
    \centering
    \includegraphics{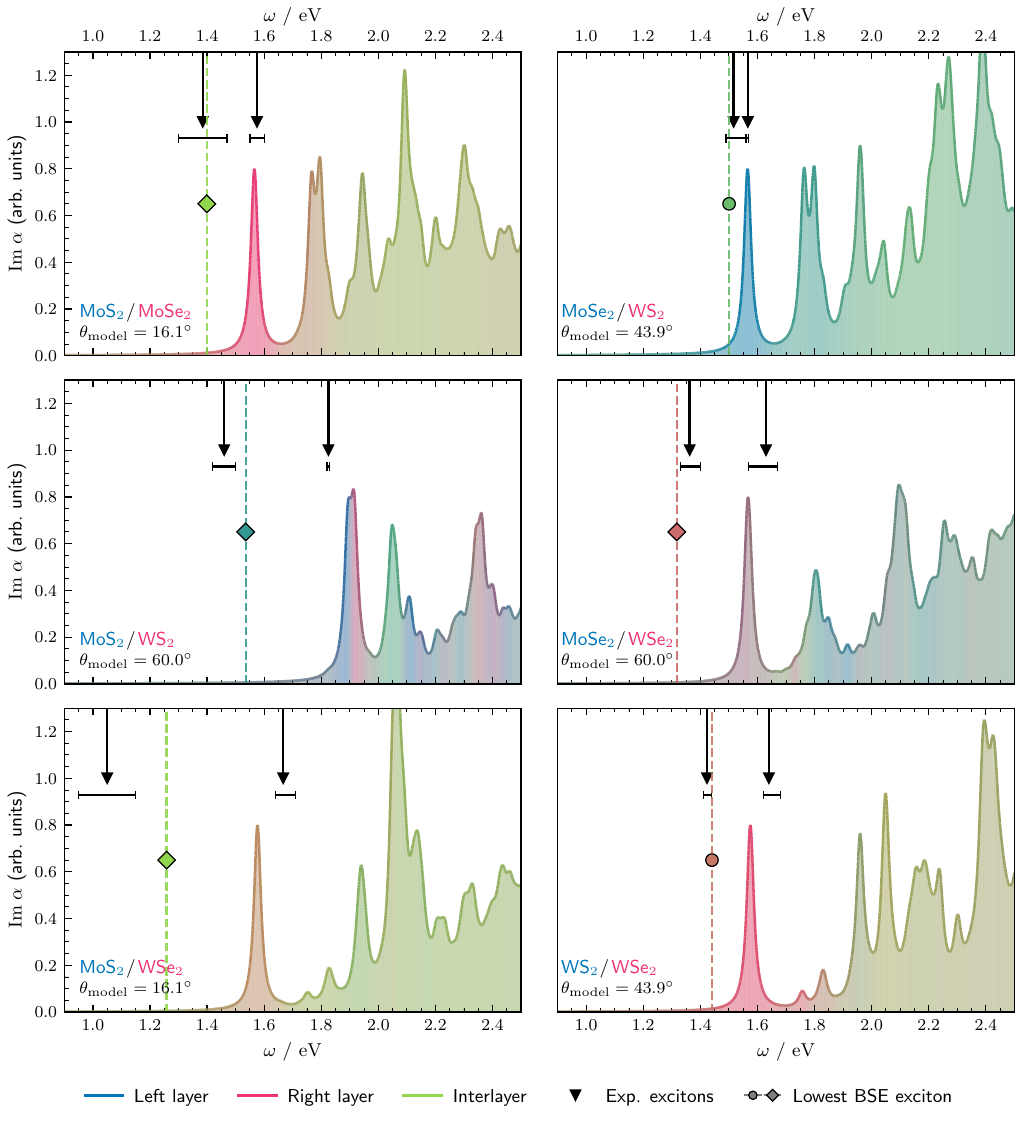}
    \caption{Absorption spectra for the six TMD heterobilayers. The spectrum is colored according to the character of the exciton, with intralayer excitons on the left (right) layer colored blue (magenta), and interlayer excitons colored green. The vertical arrows show the experimentally determined exciton excitation energies, as shown in Table \ref{TableEXP}. The variation in the experimental measurements is indicated with a horizontal line below the arrow. The lowest BSE exciton is indicated with a dashed line marked with a circle if the exciton is momentum-direct, and a diamond if it is momentum-indirect. The color of the line indicates the layer-resolved character of the exciton.}
    \label{fig:mqehbse-spectra}
\end{figure*}

For the bilayer WS$_2$/WSe$_2$, our model structure has a twist angle of $43.9^\circ$. This is very different from the lattice matched configurations in the experimental data set, yet the predicted positions of $X_0^{\mathrm{d}}$ and $X_0^{\mathrm{b}}$ are in very good agreement with the experiments - again within 50 meV of the experimental range. To justify the comparison, despite the difference in twist angle, we note from Figure \ref{fig:mqehbse-spectra} that both $X_0^{\mathrm{d}}$ and $X_0^{\mathrm{b}}$ are momentum-direct excitons. Analysis shows that they are comprised of electron-hole transitions around the $K$ and $K'$ points in the BZ of the two layers. In addition, both excitons are mainly localized in the WSe$_2$ layer: $X_0^{\mathrm{b}}$ is purely intralayer while $X_0^{\mathrm{d}}$ is mainly intralayer with a small interlayer component. Since the TMD conduction and valence bands at $K$ and $K'$ exhibit only weak interlayer hybridization, it is likely that the excitons are relatively insensitive to the twist angle. Under this assumption (which is substantiated by the example below), it is justified to compare the calculated exciton energies to the experiments.

\begin{figure*}
    \centering
    \includegraphics{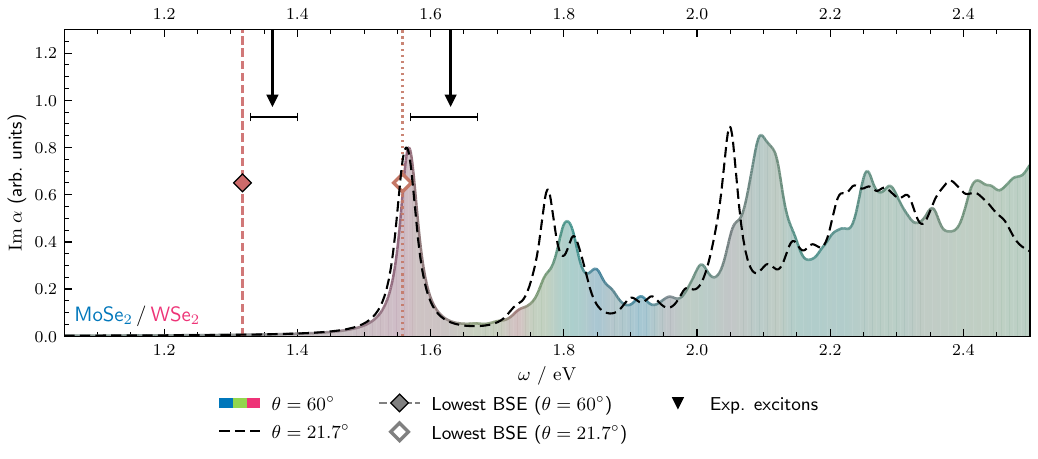}
    \caption{Calculated absorption spectra of MoSe$_2$/WSe$_2$ in two stackings: lattice-matched ($60^\circ$ twist) and twisted by $21.7^\circ$.
    The bright low-energy peak is essentially unchanged, but the low-energy indirect feature near $1.35$\,eV present in the lattice-matched structure is absent in the twisted one, reflecting twist-driven suppression of interlayer hybridization. This is consistent with the PL spectra reported in Ref.~\cite{nayak2017probing}, where a peak at 1.35 eV was observed for twist angles of 0$^\circ$ and 60$^\circ$, but which disappeared for twist angles larger than 10$^\circ$.}
    \label{fig:placeholder}
\end{figure*}

As a specific example of twist angle dependence, we consider 
MoSe$_2$/WSe$_2$ in two different twist configurations. Figure \ref{fig:placeholder} shows the calculated absorption spectrum for twist angles 60$^\circ$ (color shaded) and 21.7$^\circ$ (dashed). The lowest exciton energy at the center-of-mass momentum corresponding to the indirect band gap, i.e. $X_0^{\mathrm{d}}$, is indicated by vertical dashed lines, with filled and open diamonds denoting the 60$^\circ$ and 21.7$^\circ$ structures, respectively. The absorption spectra, including the first peak corresponding to the lowest optically active exciton, $X_0^{\mathrm{b}}$, show only weak dependence on twist angle. This can be explained from the predominantly intralayer nature of this exciton and the fact that it is comprised of transitions close to the $K$-point, which exhibit weak interlayer hybridization.  

In contrast to the weak twist angle-dependence of the bright exciton, the momentum-indirect exciton, $X_0^{\mathrm{d}}$, blue shifts by 0.25 eV (from 1.30 eV to 1.55 eV) upon twisting. This is in good agreement with photoluminescence (PL) experiments, which show that the indirect exciton observed at 1.35 eV in lattice-matched bilayers ($0^\circ$ and $60^\circ$), disappears from this energy range for intermediate twist angles\cite{nayak2017probing}. 

\begin{figure}
    \centering
    \includegraphics{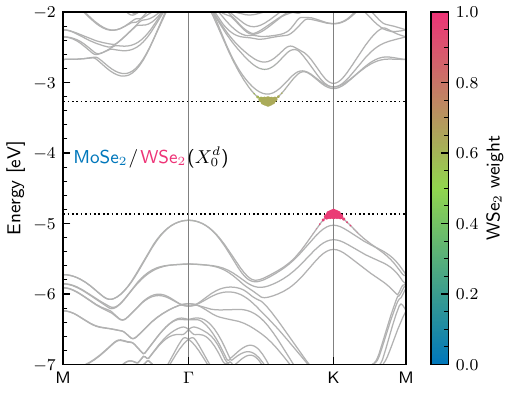}
    \caption{
    Electron-hole pair weights for the lowest exciton ($X_0^{\mathrm{d}}$) in lattice-matched MoSe$_2$/WSe$_2$ ($\theta=60^\circ$). 
    The color of the circles indicate the projection of the band onto each of the two layers.
    The hole is localized in the WSe$_2$ layer, while the electron is hybridized with significant weight on both layers. }
    \label{fig:analysis}
\end{figure}

To understand the strong twist angle-dependence of $X_0^{\mathrm{d}}$ in bilayer MoSe$_2$/WSe$_2$, we analyse the structure of its wave function. Figure \ref{fig:analysis} shows the distribution of the weights of the electron-hole pair basis in the BZ of the lattice-matched bilayer ($\theta=60^\circ$). The color of the weights indicate the projection of the electron and hole wave functions on the WSe$_2$ layer. It follows that $X_0^{\mathrm{d}}$ is a momentum-indirect, mixed intralayer/interlayer exciton with the hole at the $K$-point in WSe$_2$ and the electron midway along the $\Gamma-K$ path (the $\Lambda$-point) and delocalized over both layers with roughly 75\% weight on WSe$_2$. We ascribe the strong twist angle-dependence to the changes in conduction band states at the $\Lambda$-point upon twisting.  

 For the remaining three systems, the experimental data set includes both lattice-matched configurations, finite and/or undetermined twist angles. Consequently, direct comparison to our theoretical predictions, which, for these three systems are all based on model moiré structures with large twist angles, is less straightforward.   

The case of MoSe$_2$/WS$_2$ is very similar to WS$_2$/WSe$_2$. Indeed both $X_0^{\mathrm{b}}$ and $X_0^{\mathrm{d}}$ are momentum-direct with electron-hole transitions located around the $K$ and $K'$ points. Due to the relative inertness of the conduction and valence band states at $K$ and $K'$ to interlayer coupling, we may assume the two excitons to be twist angle-insensitive. This is supported  by the weak variation in the experimental exciton energies, despite the data set covering a range of angles between $0^\circ$ and $60^\circ$. The comparison between the experimental and predicted exciton energies, which is excellent, is thus well justified.  

For MoS$_2$/MoSe$_2$ the experimental data includes a structure with twist angle of $1^\circ$ and a structure with an undetermined twist angle. In this case, the predicted energies of both $X_0^{\mathrm{b}}$ and $X_0^{\mathrm{d}}$ fall in between the two experimental values and within 100 meV of both.  

We finally, turn to the case of MoS$_2$/WSe$_2$, which shows the largest deviation between calculations and experiments. The experimental data includes twist angles from $0^\circ$ to $9^\circ$ and shows that the energy of both 
$X_0^\mathrm{b}$ and $X_0^\mathrm{d}$ increase with twist angle. We note in passing that this trend is consistent with the behavior discussed above for MoSe$_2$/WSe$_2$, where we found that the lowest indirect exciton blue shifts by 0.25 eV when the twist angle is decreased from the lattice-matched structure at 60$^\circ$ to 21.7$^\circ$. Since our model structure for MoS$_2$/WSe$_2$ has a twist angle of $16.1^\circ$, we find it more appropriate to compare to the largest experimental energies. This leads to a deviation from experiments of slightly below (above) 100 meV for $X_0^\mathrm{b}$ ($X_0^\mathrm{d}$).


\section{Conclusion}
We have introduced the microscopic Quantum Electrostatic Heterostructure (mQEH) method for first-principles modeling of dielectric screening and excitons in van der Waals heterostructures. By representing induced densities and potentials in a hierarchical, layer-resolved dielectric basis, the method provides a systematically improvable description of microscopic screening without relying on geometric cutoffs or effective layer-thickness parameters. Combined with a layer-projected Bethe--Salpeter equation and LAPS quasiparticle band structures, the resulting mQEH-BSE framework enables quantitative calculations of exciton energies and optical spectra in heterostructures that are difficult or impractical to treat with conventional BSE implementations. Benchmark calculations for bulk MoS$_2$ and lattice-matched MoS$_2$/WS$_2$ show that the approach accurately reproduces both microscopic screening and conventional BSE spectra using a compact dielectric basis.

We applied the method to homobilayer MoS$_2$ and six TMD heterobilayers. For bilayer MoS$_2$, the calculated absorption spectrum reproduces the experimentally observed A exciton, while finite-momentum mQEH-BSE calculations capture the low-energy indirect excitons arising from the indirect bilayer band gap. For the heterobilayers, the calculated lowest optically active excitons agree very well with experimental values across all six material combinations. The method also resolves the layer character of the excitons, distinguishing intralayer, interlayer, and mixed intralayer/interlayer states. For both the lowest (possibly momentum-indirect) exciton and the lowest bright exciton, the agreement with experiment is generally better than 0.1 eV.

Overall, the mQEH-BSE framework provides an efficient and quantitatively accurate route to excitons in realistic van der Waals heterostructures. Its layer-resolved formulation makes it particularly well suited for lattice-mismatched and twisted systems, opening the door to systematic first-principles screening and design of optoelectronic properties in complex 2D material stacks.

\section{Acknowledgement}
The authors acknowledge funding from the Villum Investigator Grant No. 37789 supported by VILLUM FONDEN and from the Novo Nordisk Foundation Data Science Research Infrastructure 2022 Grant:  A high-performance computing infrastructure for data-driven research on sustainable energy materials, Grant no. NNF22OC0078009.

\appendix



\section{Layer-resolved operator representation}
\label{appendix:mqeh-basis}

In this appendix, we provide further details on the representation of the density response function and Coulomb operator in the layer-resolved basis introduced in Section~\ref{sec:theory-mqeh}, and on the solution of the interlayer Dyson equation~\eqref{eq:dyson-interlayer}.

\subsection{Operator representation and the matrix Dyson equation}

We write the density response function $\chi$ (and similarly $\wt{\chi}$) in the operator representation
\begin{equation}
\label{eq:appendix-chi-representation}
\chi(\q, \omega) = \sum_{ij} \ket{\rho_i(\q)} \chi_{ij}(\q, \omega) \bra{\rho_j(\q)},
\end{equation}
where $i = (\ell, \alpha)$ and $j = (\ell', \beta)$ are combined layer and basis function indices.
The density induced by an external potential $V_\text{ext}(\q, \omega, z)$ is then
\begin{equation}
\label{eq:appendix-rho-ind}
\begin{aligned}
\rho_\text{ind} & (\q, \omega, z) 
 = \sum_{ij} \Big[ \\ 
 &\rho_i(\q, z)\, \chi_{ij}(\q, \omega) \int \rho_j^*(\q, z')\, V_\text{ext}(\q, \omega, z') \dd z' \Big].
\end{aligned}
\end{equation}
By the duality condition~\eqref{eq:orthonormality-rho-phi}, the integrals $d^\ell_\alpha \equiv \braket{\rho^\ell_\alpha}{V_\text{ext}}$ appearing in Eq.~\eqref{eq:appendix-rho-ind} may be interpreted as the expansion coefficients of $V_\text{ext}$ in a \emph{local} expansion
\begin{equation}
\label{eq:appendix-Vext-local}
V_\text{ext}(\q, \omega, z) \approx \sum_\alpha d^\ell_\alpha(\q, \omega)\, \phi^\ell_\alpha(\q, z),
\end{equation}
valid in the vicinity of layer $\ell$.
Notably, Eq.~\eqref{eq:appendix-Vext-local} does not imply a globally valid representation of $V_\text{ext}$ in terms of the potential basis functions; rather, it provides the coefficients needed to determine the induced density on layer $\ell$ via Eq.~\eqref{eq:appendix-rho-ind}.

We define the Coulomb matrix elements by
\begin{equation}
\label{eq:appendix-coulomb_matrix}
V_{ij}(\q) = \mel{\rho_i(\q)}{V(q)}{\rho_j(\q)},
\end{equation}
which may be interpreted as the electrostatic interaction energy between the charge distributions $\ket{\rho_i(\q)}$ and $\ket{\rho_j(\q)}$.
In practice, the matrix elements are obtained by solving the Poisson equation for each density basis function $\ket{\rho_j(\q)}$ and computing the overlap of the resulting potential with the basis function $\ket{\rho_i(\q)}$.

Inserting the operator representation~\eqref{eq:appendix-chi-representation} into the interlayer Dyson equation~\eqref{eq:dyson-interlayer}, $\chi = \wt{\chi} + \wt{\chi} V^I \chi$, yields the matrix form of the Dyson equation
\begin{equation}
\label{eq:appendix-dyson-matrix}
\chi_{ij}(\q, \omega) = \wt{\chi}_{ij}(\q, \omega) + \sum_{nm} \wt{\chi}_{in}(\q, \omega)\, V^I_{nm}(\q)\, \chi_{mj}(\q, \omega),
\end{equation}
where $V^I_{ij}(\q) = V_{ij}(\q)(1-\delta_{\ell\ell'})$ is the interlayer part of the Coulomb matrix.
The uncoupled response $\wt{\chi}$ is block-diagonal in the layer index by construction, since it contains no interlayer coupling. Its matrix elements are obtained from the intralayer response of each isolated monolayer, as described in Section~\ref{appendix:finite-freq} below. The Coulomb matrix $V^I$, on the other hand, couples different layers, so the solution $\chi$ of Eq.~\eqref{eq:appendix-dyson-matrix} is not block-diagonal, reflecting the physical effect of interlayer screening.
The dimension of the matrix equation is $\sum_\ell N_\ell$, which is typically small (around 10--20 for the bilayers considered here), making the solution computationally inexpensive.

\subsection{Mixed-space representation and in-plane diagonal approximation}
\label{appendix:mixed-space}

In practice, $\wt{\chi}^\ell$ is computed in a plane-wave basis using the GPAW~\cite{mortensen2024gpaw} code.
This plane-wave basis is related to the real-space representation by
\begin{equation}
    \wt{\chi}^\ell(\q, \r, \r')
    = \frac{1}{\Omega_\ell} \sum_{\G \G'}
    e^{i \left(\G + \q \right)\cdot \r}
    \wt{\chi}^\ell_{\G \G'}(\q)
    e^{-i \left(\G' + \q \right) \cdot \r'},
\end{equation}
where $\Omega_\ell$ is the volume of the computational unit cell of monolayer $\ell$, and
\begin{equation}
    \wt{\chi}^\ell(\r, \r')
    = \frac{1}{N_\q}\sum_{\q \in \text{BZ}} \wt{\chi}^\ell(\q, \r, \r').
\end{equation}
{}
To avoid spurious interactions between repeated images of the monolayer due to the periodic boundary conditions imposed by the plane-wave basis, we use a truncated Coulomb interaction~\cite{ismail-beigi2006truncation} in the solution of Eq.~\eqref{eq:dyson-intralayer}.

To connect to the layer-resolved basis, we separate in-plane and out-of-plane coordinates, writing $\r = (\r_\parallel, z)$ and $\G = (\G_\parallel, G_z)$, and introduce the mixed-space representation
\begin{equation}
    \wt{\chi}_{\G_\parallel, \G_\parallel'}^\ell(\q, z, z')
    = \frac{1}{L_\ell} \sum_{G_z G'_z}
    e^{i G_z z}
    \wt{\chi}^\ell_{\G \G'}(\q)
    e^{-i G_z' z'},
\end{equation}
where $L_\ell$ is the out-of-plane length of the computational unit cell of monolayer $\ell$.
In a homogeneous 2D system,
$\wt{\chi}$ is diagonal in $\G_\parallel$ and $\G_\parallel'$. 
The off-diagonal elements describe higher-order in-plane Fourier components of the induced charge density, representing processes where momentum is transferred between the external field and the lattice via scattering from the periodic crystal potential. 
For heterostructures with large Moir\'{e} supercells, the number of in-plane reciprocal vectors becomes extremely large, rendering the full treatment impractical.
We therefore neglect the off-diagonal terms and adopt the approximation
\begin{equation}
\label{eq:appendix-chi-diagonal-in-G_parallel}
\wt{\chi}_{\G_\parallel, \G_\parallel'}^\ell \approx \wt{\chi}_{\G_\parallel, \G_\parallel}^\ell \delta_{\G_\parallel, \G_\parallel'}.
\end{equation}
This approximation neglects only \emph{interlayer} local field effects, in which electrons are scattered to a different in-plane momentum by charge fluctuations in a neighboring layer.
Intralayer local field effects are fully retained through the solution of Eq.~\eqref{eq:dyson-intralayer} in the full plane-wave basis.
Moreover, because the interlayer Dyson equation~\eqref{eq:appendix-dyson-matrix} is solved independently at each $\q + \G_\parallel$, the approximation~\eqref{eq:appendix-chi-diagonal-in-G_parallel} enables the treatment of non-commensurate and large Moir\'{e} supercell heterostructures at no additional computational cost compared to commensurate bilayers.

\subsection{Representation of \texorpdfstring{$\wt{\chi}^\ell$}{chi-tilde} at finite frequency and long wavelengths}
\label{appendix:finite-freq}

The shape of the eigenfunctions $\phi^\ell_\alpha(\q, z)$ and $\rho^\ell_\alpha(\q, z)$ changes only weakly with frequency $\omega$. To avoid storing nearly identical basis functions at every frequency point, we diagonalize $V(\q)\wt{\chi}^\ell(\q, \omega\!=\!0)$ only at $\omega = 0$ and use the resulting eigenfunctions as the density and potential basis at all frequencies.
At finite frequency, $\wt{\chi}^\ell(\q, \omega)$ is therefore not strictly diagonal in the basis function indices, but still permits the general representation~\eqref{eq:appendix-chi-representation}.
The coefficients $\wt{\chi}^\ell_{\alpha \beta}(\q, \omega)$ are obtained by a constrained least-squares fit of the induced density
\begin{equation}
\rho^\text{ind}_{\ell \beta}(\q, \omega, z) = \int \dd z'\, \wt{\chi}^\ell (\q, \omega, z, z')\, \phi^{\ell}_{\beta} (\q, z')
\end{equation}
in terms of the density basis functions $\ket{\rho_{\alpha}^{\ell}(\q)}$,
subject to the constraint that the integral of the induced density is preserved exactly,
\begin{equation}
\label{eq:appendix-least-squares-constraint}
    \int \dd z\, \rho^{\text{ind}}_{\ell \beta}(\q, \omega, z) 
    = 
    \sum_\alpha \wt{\chi}^\ell_{\alpha \beta}(\q, \omega) \int \dd z\, \rho^\ell_\alpha(\q, z). 
\end{equation}
This constraint is particularly important in the $q \to 0$ limit, where the Coulomb interaction diverges as $V(q) \sim 1/q$. This divergence is canceled by the corresponding asymptotic behavior of $\wt{\chi}^\ell$, but when the basis is truncated to a finite number $N_\ell$ of functions, a straightforward projection of $\rho^\text{ind}$ may introduce small residual errors that violate charge conservation and produce spurious divergences in the induced potentials.
For finite $q$, charge neutrality of the density basis functions follows from the assumed in-plane variation $e^{i\q \cdot \r_\parallel}$. In the $q \to 0$ limit, charge neutrality must instead be enforced by requiring $\int \rho^\text{ind}(\q,\omega, z)\, \dd z = 0$, making the constraint~\eqref{eq:appendix-least-squares-constraint} essential in this regime.

In addition, the eigenfunctions $\phi^\ell_\alpha(q,z)$ become increasingly oscillatory in $z$ as $q \to 0$, reflecting the ill-conditioned nature of the Coulomb kernel in this limit. These oscillatory functions become poorly suited for representing smooth external potentials. We therefore introduce a momentum threshold $q_{\text{threshold}} = 0.1\,\text{Bohr}^{-1}$; for $q < q_{\text{threshold}}$, the basis functions are fixed to those obtained at $q = q_{\text{threshold}}$, providing a more stable representation in the long-wavelength limit.

We emphasize that the mathematical validity of the formalism is independent of the specific choice of basis, provided that the density and potential basis functions satisfy the required duality relation~\eqref{eq:orthonormality-rho-phi} within each layer. The basis can always be made complete by including a sufficient number of basis functions.
\section{Convergence with k-points}

  \begin{figure}
      \centering
      \includegraphics{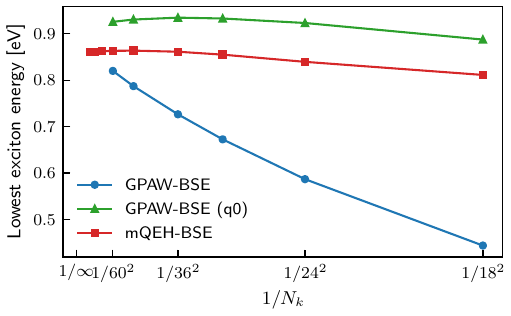}
      \caption{Convergence of the lowest BSE eigenvalue with the number of
      k-points, $N_k$, for lattice-matched MoS${}_2$-WS${}_2$. The mQEH-BSE
      method is compared to the GPAW-BSE implementation with and without the
      $q_0$-correction of Ref.~\cite{rasmussen2016efficient}. All calculations
      are performed in PW mode without spin--orbit coupling, on top of PBE
      eigenvalues without a scissors shift (hence the small exciton energies).
      The GPAW-BSE calculations extend to a maximum of $60\times60$ k-points,
      beyond which they become computationally unfeasible, while the mQEH-BSE
      calculations extend to $96\times96$ k-points.}
      \label{fig:bse-convergence}
  \end{figure}

  Figure~\ref{fig:bse-convergence} shows the convergence of the lowest
  $\vb{q}=0$ BSE eigenvalue with respect to the density of the k-point grid,
  comparing the mQEH-BSE method to the BSE implementation in
  GPAW~\cite{mortensen2024gpaw}, both with and without the so-called
  $q_0$-correction~\cite{rasmussen2016efficient}, which applies an analytical
  correction to the $\vb{q}=0$ component of the screened interaction.

  Without the $q_0$-correction, GPAW-BSE converges very slowly and remains far
  from its asymptotic value even at the densest grid we could afford
  ($60\times60$). With the correction, the convergence rate becomes comparable
  to that of mQEH-BSE. 
  This identifies the origin of the slow convergence as due to the matrix elements of the BSE kernel -- in
  particular the direct, screened part -- not being converged at coarse k-point samplings.
  The mQEH-BSE method
  avoids this problem, since the screened interaction is
  evaluated from dielectric building blocks obtained from monolayer
  calculations at high k-point densities, independently of the grid used for
  the BSE Hamiltonian itself.

  Beyond this initial regime, where the exciton energy increases with k-point
  density, both the $q_0$-corrected GPAW-BSE and the mQEH-BSE energies pass
  through a maximum and then decrease slowly with further grid refinement,
  consistent with earlier findings for bulk
  materials~\cite{alvertis2023importance}. This slow decrease reflects a
  different regime for convergence: it is no longer the screened interaction
  that limits the accuracy, but rather the fine k-point sampling needed to
  resolve the exciton wavefunction in reciprocal space.

  From the available data it is not possible to extrapolate to the infinite density limit with sufficient accuracy to determine how well the three methods would agree in this limit.
\section{Computational details} 
\label{appendix:computational-details} 
 
All atomic structures were relaxed with the PBE exchange--correlation functional~\cite{perdew1996generalized} including the D3 dispersion correction~\cite{grimme2010consistent}.
The lattice-mismatched heterobilayers were modeled by near-coincidence moir\'e supercells built from the relaxed monolayer lattices.
The residual in-plane strain required to make each bilayer commensurate is listed in Table~\ref{tab:moire-cells}. 
 
Ground-state calculations were performed with the GPAW code~\cite{mortensen2024gpaw,enkovaara2010electronic} using PBE in LCAO mode with a 
double-$\zeta$ polarized (DZP) basis set~\cite{larsen2009localized}.
The quasiparticle band structure and the self-consistent interlayer band alignment were obtained by adding the LAPS operator [Eq.~\eqref{eq:LAPS-operator}] to the Kohn--Sham Hamiltonian~\cite{leon2025laps}, with the layer-resolved scissors shifts fixed by the G$_0$W$_0\Gamma_0$ monolayer band-edge targets of Table~\ref{tab:GWmonolayers} (obtained as described in Refs.~\cite{leon2025laps,schmidt2017simple}) together with the G$\Delta$W image-charge correction~\cite{winther2017band}.
Spin--orbit coupling was included throughout, except in the technical benchmarks of Figs.~\ref{fig:bse-mqeh-gpaw} and~\ref{fig:bse-convergence}. 
 
The mQEH dielectric building blocks were computed within the RPA from a plane-wave ground state with an $800$~eV plane-wave cutoff, on a $128\times128$ Monkhorst--Pack $k$-point grid, including $400$ bands in the response function, a $250$~eV cutoff for the dielectric matrix, and a broadening of $\eta = 50$~meV.
A truncated Coulomb interaction~\cite{ismail-beigi2006truncation} was used to eliminate interactions between periodic images of the monolayer. In each layer we kept $12$ dielectric basis functions, and used the long-wavelength threshold $q_\mathrm{threshold}=0.1~\mathrm{Bohr}^{-1}$ (defined in Appendix~\ref{appendix:mqeh-basis}). 
 
The mQEH-BSE Hamiltonian was set up on $\Gamma$-centered Monkhorst--Pack $k$-point grids using a $150$~eV plane-wave cutoff for the pair densities. 
For the commensurate systems, we used a $60\times60$ k-point grid and four valence and four conduction bands. 
For the $21.7^\circ$ MoSe$_2$/WSe$_2$ cell, we used a $15\times15$ grid with $12$ valence and $12$ conduction bands. For the four remaining heterobilayers ($75$ atoms), we used a $12\times12$ grid with $22$ valence and $22$ conduction bands. 
Momentum-indirect excitons were obtained by solving the BSE at the finite center-of-mass momentum $Q$ corresponding to the indirect band gap of each system (e.g.\ $Q=1/3$ along $\Gamma$--K for the homobilayer). 
A broadening of $\eta = 15$~meV was used in the calculation of absorption spectra. 

\section{Experimental data}
In this section, we present tables of the experimentally determined values of exciton energies for all structures studied.
\setlength{\tabcolsep}{1.2em}
\begin{table*}[h!]
  \begin{center}
    \begin{tabular}{cccccc}
      \hline
      \hline
      \rule{0pt}{2.5ex}
      System & $E^A$ & $E^B$ & Substrate & Temperature & Technique \\
      \hline
      \rule{0pt}{2.5ex}
      MoS$_2$\cite{mak2010atomically}    & 1.85  & 1.98          & SiO$_2$ & Room temp. & Refl./PL \\
      MoS$_2$\cite{cadiz2017excitonic}   & 1.93--1.95 & --       & hBN     & 4 K        & PL \\
      MoS$_2$\cite{li2014measurement}    & 1.86  & 2.02         & fused silica & Room temp. & Refl. \\
      MoS$_2$\cite{goryca2019revealing}  & 1.939 & 
      $\sim$2.09   & hBN     & 4 K        & Magneto-abs. \\
      \hline
      \rule{0pt}{2.5ex}
      WS$_2$\cite{zhao2013evolution}     & 2.00  & 2.39          & SiO$_2$ & Room temp. & Diff. refl./PL \\
      WS$_2$\cite{chernikov2014exciton}  & 2.09  & --            & SiO$_2$ & 5 K        & Refl. contrast \\
      WS$_2$\cite{li2014measurement}    & 2.02  & 2.41         & fused silica & Room temp. & Refl. \\
      WS$_2$\cite{goryca2019revealing}   & 2.058 & $\sim$2.45    & hBN     & 4 K        & Magneto-abs. \\
      \hline
      \rule{0pt}{2.5ex}
      MoSe$_2$\cite{ross2013electrical}   & 1.659 & $\sim$1.85           & SiO$_2$ & 20 K       & PL \\
      MoSe$_2$\cite{li2014measurement}    & 1.55  & 1.74         & fused silica & Room temp. & Refl. \\
      MoSe$_2$\cite{cadiz2017excitonic}   & 1.63 & --       & hBN     & 4 K        & PL \\
      MoSe$_2$\cite{goryca2019revealing}  & 1.643 & $\sim$1.84      & hBN     & 4 K        & Magneto-abs. \\
      \hline
      \rule{0pt}{2.5ex}
      WSe$_2$\cite{zhao2013evolution}             & 1.67   & 2.03--2.08 & SiO$_2$ & Room temp. & Diff. refl./PL \\
      WSe$_2$\cite{li2014measurement}    & 1.65  & --         & fused silica & Room temp. & Refl. \\
      WSe$_2$\cite{stier2018magnetooptics}        & 1.723  & --   & hBN     & 4 K        & Magneto-abs. \\
      WSe$_2$\cite{liu2019magnetophotoluminescence} & 1.712 & -- & hBN     & 4 K        & PL \\
      \hline
    \end{tabular}
    \caption{Experimentally measured A and B exciton energies in monolayer TMDs. All energies are in eV. These values are used for the experimental arrows in Fig.~\ref{fig:monolayer-spectra}.}
    \label{TableEXP_ML}
  \end{center}
\end{table*}
\begin{table*}[h!]
  \begin{center}
    \begin{tabular}{ccccccc}
      \hline
      \hline
      \rule{0pt}{2.5ex}
      System & $E^{\mathrm{indir}}$ & $E^A$ & $E^{\mathrm{IL}}$ & Substrate & Temperature & Technique \\
      \hline
      \rule{0pt}{2.5ex}
      H-MoS$_2$\cite{mak2010atomically}         & 1.59   & 1.86   & --     & fused silica/suspended       & Room temp.   & PL/Abs. \\
      H-MoS$_2$\cite{conley2013bandgap}          & 1.53  & 1.82   & --     & polycarbonate      & Room temp.   & PL \\
      H-MoS$_2$ \cite{grzeszczyk2021optical} & 1.51 & 1.93   & 2.00 & hBN & 5 K & PL \\
      H-MoS$_2$ \cite{peimyoo2021electrical} & -- & 1.93  & 1.99 & hBN & 4 K & Refl. contrast \\
      \hline
    \end{tabular}
    \caption{Experimentally measured exciton energies in MoS$_2$ homobilayers in H stacking. $E^A$ is the lowest K-point intralayer exciton (the A:1s exciton). $E^{\mathrm{indir}}$ is the energy of the lowest momentum-indirect exciton corresponding to the lowest feature in the photoluminescence spectrum. $E^{\mathrm{IL}}$ has been described as a mixed intralayer/interlayer exciton -- a hybrid of the intralayer B exciton and pure interlayer B exciton\cite{peimyoo2021electrical}. All energies are in eV. 
    The values in this table are used for the experimental arrows in Fig.~\ref{fig:MoS2-bilayer-spectra}.}
    
    \label{TableEXP_2L_MoS2}
  \end{center}
\end{table*}
\begin{table*}[h!]
  \begin{center}
     \begin{tabular}{cccccc}
      \hline
      \hline
      \rule{0pt}{2.5ex}
      System & $E_0^{\mathrm{d}}$ & $E_0^{\mathrm{b}}$ & Twist angle & Substrate & Temperature \\
      \hline
      \rule{0pt}{2.5ex}
      MoS$_2$/MoSe$_2$\cite{alexeev2024nature} & 1.30 & 1.55 & $1^\circ$ & BN & Room temp. \\
      MoS$_2$/MoSe$_2$\cite{mouri2017thermal} & 1.47 & 1.60 & N/A & SiO$_2$ & 5 K \\
      \hline
      \rule{0pt}{2.5ex}
      MoS$_2$/WS$_2$\cite{gong2014vertical} & 1.42 & 1.82 & $60^\circ$ & SiO$_2$ & Room temp. \\
      MoS$_2$/WS$_2$\cite{heo2015rotation} & 1.5 & 1.83 & $0^\circ$ & SiO$_2$ & Room temp. \\
      \hline
      \rule{0pt}{2.5ex}
      MoS$_2$/WSe$_2$\cite{karni2019infrared} & 1.05 & 1.65 & $0^\circ$ & Quartz & 20 K \\
      MoS$_2$/WSe$_2$\cite{lin2024moire} & 0.95 -- 1.15 & 1.64 -- 1.71 & $0.5^\circ$ -- $9^\circ$ & PDMS & Room temp. \\
      \hline
      \rule{0pt}{2.5ex}
      MoSe$_2$/WS$_2$\cite{alexeev2019resonantly} & 1.50--1.56 & 1.57 & $0^\circ$ -- $60^\circ$ & SiO$_2$ & Room temp. \\
      MoSe$_2$/WS$_2$\cite{alexeev2017imaging} & 1.49 & 1.56 & $1^\circ$ & SiO$_2$ & Room temp. \\
      \hline
      \rule{0pt}{2.5ex}
      MoSe$_2$/WSe$_2$\cite{ciarrocchi2019polarization} & 1.40 & 1.62 & $\leq 1^\circ$ & BN & 4.2 K \\
      MoSe$_2$/WSe$_2$\cite{jiang2018microsecond} & 1.34 & 1.67 & $0^\circ$ & SiO$_2$ & 2.3 K \\
      MoSe$_2$/WSe$_2$\cite{miller2017long} & 1.33 & 1.66 & $60^\circ$ & Schott borofloat\textsuperscript{\textregistered}33 & 3 K \\
      MoSe$_2$/WSe$_2$\cite{wilson2017determination} & 1.38 & 1.57 & $<1^\circ$ & Graphite & Room temp. \\
      \hline
      \rule{0pt}{2.5ex}
      WS$_2$/WSe$_2$\cite{jin2019observation} & 1.41 & 1.68 & $0.5^\circ$ & BN & 10 K \\
      WS$_2$/WSe$_2$\cite{rossi2024anomalous} & 1.42 & 1.62 & $0^\circ$ & BN & 30 K \\
      WS$_2$/WSe$_2$\cite{rossi2024anomalous} & 1.44 & 1.62 & $60^\circ$ & BN & 30 K \\
    \end{tabular}
    \caption{Overview of experimentally measured energies of the lowest bright
    exciton ($X_0^{\mathrm{b}}$), the lower of the two intralayer A excitons) and
    the lowest dark exciton ($X_0^{\mathrm{d}}$), interlayer- or momentum-indirect) in TMD heterobilayers, together with the interlayer twist angle, the
    substrate on which the bilayer was placed, and the temperature at which the
    measurement was conducted. Twist angles marked N/A were not controlled or
    measured; a range of angles indicates a twist-angle series, with the quoted
    energies spanning the same series. All energies are in eV. All energies represent
    photoluminescence peak positions, except the $X_0^{\mathrm{b}}$ values of
    Refs.~\cite{mouri2017thermal} (photoluminescence excitation),
    \cite{lin2024moire} and \cite{jin2019observation} (reflection contrast).
    These values are used for the experimental arrows in Fig.~\ref{fig:mqehbse-spectra}.}
    \label{TableEXP}
  \end{center}
\end{table*}

\bibliography{bibliography} 
\end{document}